\documentclass[twocolumn,superscriptaddress,english,showpacs,longbibliography]{revtex4-2}
\usepackage[colorlinks=true,urlcolor=blue,citecolor=blue,linkcolor=blue]{hyperref} 

\usepackage{amsmath}
\usepackage{graphicx}
\usepackage{textcomp}
\usepackage{bm}
\usepackage{color}
\usepackage{amssymb}
\usepackage{graphicx}
\usepackage{color}
\usepackage{mathrsfs}
\usepackage{float}
\usepackage{indentfirst}
\usepackage{txfonts}
\usepackage{algorithm}  
\usepackage{algpseudocode}  
\usepackage{balance}
\usepackage[keeplastbox]{flushend}

\newcommand{\Eq}[1]{Eq.~(\ref{#1})}
\newcommand{\Fig}[1]{Fig.~\ref{#1}}
\newcommand{\App}[1]{Appendix~\ref{#1}}

\begin{document}


\title{Tropical Tensor Network for Ground States of Spin Glasses}

\author{Jin-Guo Liu}
\email{cacate0129@gmail.com}
\affiliation{
Beijing National Lab for Condensed Matter Physics and Institute of Physics, Chinese Academy of Sciences, Beijing 100190, China
}
\affiliation{
Harvard University, Cambridge, Massachusetts 02138, United States
}
\affiliation{
QuEra Computing Inc., Boston, Massachusetts 02143, United States
}

\author{Lei Wang}
\email{wanglei@iphy.ac.cn}
\affiliation{
Institute of Physics, Chinese Academy of Sciences, Beijing 100190, China
}
\affiliation{
Songshan Lake Materials Laboratory, Dongguan, Guangdong 523808, China
}

\author{Pan Zhang}
\email{panzhang@itp.ac.cn}
\affiliation{
 CAS Key Laboratory for Theoretical Physics, Institute of Theoretical Physics, Chinese Academy of Sciences, Beijing 100190, China}
\affiliation{School of Fundamental Physics and Mathematical Sciences, Hangzhou Institute for Advanced Study, UCAS, Hangzhou 310024, China}
\affiliation{International Centre for Theoretical Physics Asia-Pacific, Beijing/Hangzhou, China}


\begin{abstract}
    We present a unified exact tensor network approach to compute the ground state energy, identify the optimal configuration, and count the number of solutions for spin glasses. The method is based on tensor networks with the \textit{Tropical Algebra} defined on the semiring of $\left(\mathbb R \cup \{-\infty\},\oplus,\odot\right)$. Contracting the tropical tensor network gives the ground state energy; differentiating through the tensor network contraction gives the ground state configuration; mixing the tropical algebra and the ordinary algebra counts the ground state degeneracy. The approach brings together the concepts from graphical models, tensor networks, differentiable programming, and quantum circuit simulation, and easily utilizes the computational power of graphical processing units (GPUs). For applications, we compute the exact ground state energy of Ising spin glasses on square lattice up to $1024$ spins, on cubic lattice up to $216$ spins, and on $3$ regular random graphs up to $220$ spins, on a single GPU; We obtain exact ground state energy of $\pm J$ Ising spin glass on the chimera graph of D-Wave quantum annealer of $512$ qubits in less than $100$ seconds and investigate the exact value of the residual entropy of $\pm J$ spin glasses on the chimera graph; Finally, we investigate ground-state energy and entropy of $3$-state Potts glasses on square lattices up to size $18\times 18$. Our approach provides baselines and benchmarks for exact algorithms for spin glasses and combinatorial optimization problems, and for evaluating heuristic algorithms and mean-field theories.
\end{abstract}

\maketitle

\section{Introduction}
Combinatorial optimization problems are fundamental to theoretical studies in statistical physics and computer science. Efficient solutions to combinatorial optimization problems are also relevant to many practical applications such as operations research and artificial intelligence. A prototypical combinatorial optimization problem is finding the ground state of the Ising spin glass with the energy function
\begin{equation}
    E(\{\sigma\}) = -\sum_{i < j }J_{ij} \sigma_i \sigma_j  - \sum_i h_i \sigma_i,
\label{eq:problem}
\end{equation}
where $\{\sigma\}\in\{\pm 1\}^N$ denotes a configuration of $N$ Ising spins. Such problem arises in broad contexts ranging from magnetic properties of dilute alloys~\cite{edwards1975theory} to probabilistic inference in graphical models~\cite{koller2009probabilistic}. 
Finding the ground state of the spin-glass is NP-hard except on some special graphs~\cite{Barahona1982}. This implies that an efficient solution to the problem is unlikely unless P = NP. Many NP problems have convenient Ising spin glass formulation~\cite{Lucas2014}. In past decades, various approaches have been applied to such a problem, including simulated annealing on classical computers~\cite{Kirkpatrick1983} and quantum annealing on manufactured quantum devices~\cite{Johnson2011}.

Besides the ground state energy and configurations, counting the number of ground-state configurations is also of interest from a physics and optimization perspective. 
The number of degeneracy characterizes the level of frustration and gives rise to residual entropy of the system at zero temperature~\cite{pauling1935structure}. For example, there can be an exponentially large number of degenerated ground states of the spin-glass such that the system exhibits finite entropy density in the thermodynamic limit. Unfortunately, counting the number of the degenerated ground state of spin glasses is \#P-complete~\cite{valiant1979complexity} which can be even harder than finding the ground state.  

In this paper, we present a unified approach to compute ground state energy, find out the ground state configuration, and count the ground state degeneracy of spin glasses exactly. The approach is based on the exact contraction of the tensor networks with tropical numbers which compute the spin-glass partition function directly in the zero-temperature limit. In the principle, the approach is not conceptually new since there can be equivalent dynamic programming or message passing formulations. It is rather a synthesis of techniques in combinatorial optimization,  graphical model, and machine learning into a unified framework in the language of tensor networks, which provides valuable insights for efficient and generic implementations. In particular, the tropical tensor network offers a general computational framework so that one can easily exploit software and hardware advances in quantum circuit simulations, automatic differentiation, and hardware accelerations. In this regard, the approach adds another example along the fruitful line of research bridging the graphical models, tensor networks, and quantum circuits~\cite{treewidth,Critch2014, Chen2017h, Han2018, Glasser, Boixo2017, Gao2018, Robeva2019, Pan2020}. 

There were previous efforts of investigating low-temperature properties of spin-glasses using approximated tensor contraction methods~\cite{Wang2014e,Rams2018a,Pan2020}. Among other things, these approaches and the related transfer matrix approach~\cite{Morgenstern1979, Cheung1983} face numerical issues at low temperatures due to the cancellation of tensor elements with exponential scales~\cite{Zhu2019b}. 
References \cite{Vanderstraeten2018a, Vanhecke2020a} investigated the residual entropy of infinite translational invariant frustrated classical spin systems by constructing tensor networks according to local rules of the ground-state manifold. More closely related to the present paper, one can employ exact tensor network contraction to count the number of solutions in the constraint
satisfaction problems~\cite{Garcia-Saez, Biamonte2015, kourtis2019fast, DeBeaudrap2020}, however, with the ground-state energy known to be zero a priori.

\section{Tropical Tensor Network}
Tropical algebra is defined by replacing the usual sum and product operators for ordinary real numbers with the max and sum operators respectively~\cite{maclagan2015introduction} 
\begin{eqnarray}
x \oplus y  = \max(x, y),\,\,\,\,\,\,\,\,\,\,\,\, \,\,\,\,
x \odot y   =  x + y. \label{eq:max-sum-alg}
\end{eqnarray}
One sees that $-\infty$ acts as zero element for the tropical number since  $-\infty \oplus x = x  $ and $-\infty \odot x = -\infty$. On the other hand, $0$ acts as the multiplicative identity since $0 \odot x = x$. The $\oplus$ and $\odot$ operators still have commutative, associative, and distributive properties. However, since there is no additive inverse, the $\oplus$ and $\odot$ and operations define a semiring over $\mathbb R \cup \{-\infty\} $. 
The semiring formulation unifies a large number of inference algorithms in the graphical models based on dynamic programming~\cite{Kschischang2001,Aji2000}. Recently, there have been efforts in combing the semiring algebra with modern deep learning frameworks with optimized tensor operations and automatic differentiation~\cite{Obermeyer2019, Rush2020}.

\begin{figure}[t]
\centering
    \includegraphics[width=\columnwidth, trim={1cm 2cm 0 0}, clip]{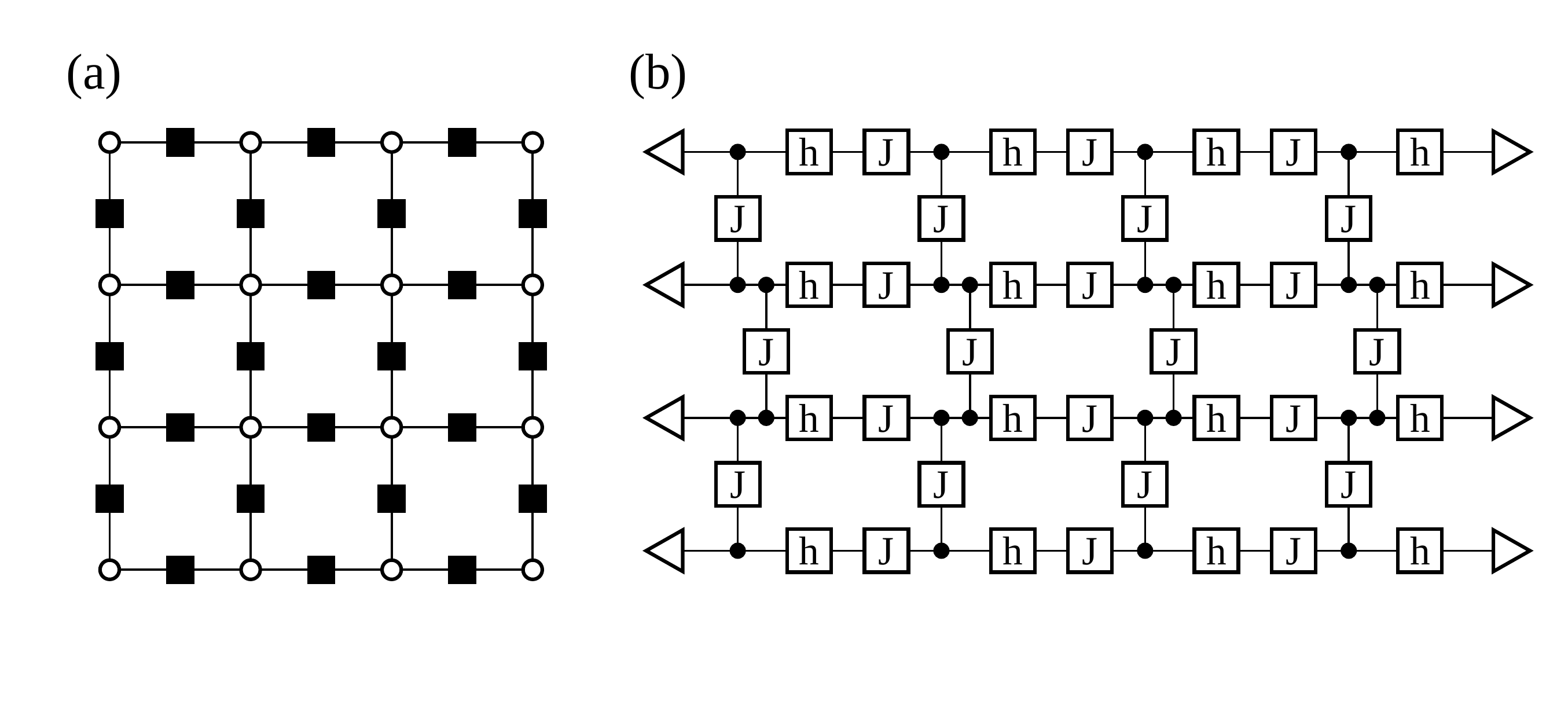}
    \caption{(a) The tensor network representation of a square lattice Ising spin glass. (b) An equivalent circuit representation used for the practical simulation. See text for the definition of the symbols.  
 \label{fig:square}}
\end{figure}

One can consider tensor networks whose elements are tropical numbers with the algebra \Eq{eq:max-sum-alg}. Since the elementary operations involved in contracting tensor networks are just sum and product, the contraction of tropical tensor networks is well defined. One can use such contraction to solve the ground state of the Ising spin glass. For example, consider the Ising spin glasses \Eq{eq:problem} defined on two-dimensional square lattice, the tropical tensor network is shown in Fig.~\ref{fig:square}(a). The tensor network representation corresponds to the factor graph of the spin-glass graphical model~\cite{Kschischang2001}.
There are $2\times 2$ tropical tensors $
    \raisebox{-2.4ex}{\includegraphics[scale=0.25, trim={0cm 0.0cm 0cm 0cm}, clip]{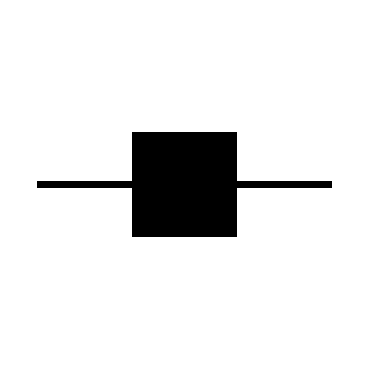}}= \left(\begin{array}{cc} J_{ij} & -J_{ij} \\  -J_{ij} & J_{ij} \end{array}\right)$ reside on the bond connecting vertices $i$ and $j$, with the tensor elements being the negative coupling energies.
The dots are diagonal tensors with $\raisebox{-4.2ex}{\includegraphics[scale=0.4, trim={0cm 0.0cm 0cm 0cm}, clip]{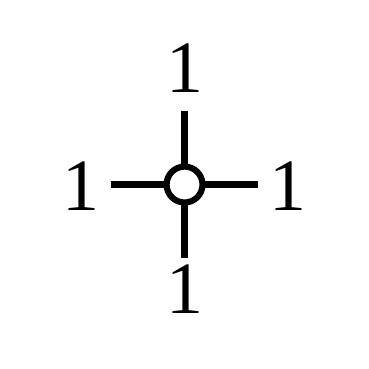}}= h_i$,  $\raisebox{-4.2ex}{\includegraphics[scale=0.4, trim={0cm 0.0cm 0cm 0cm}, clip]{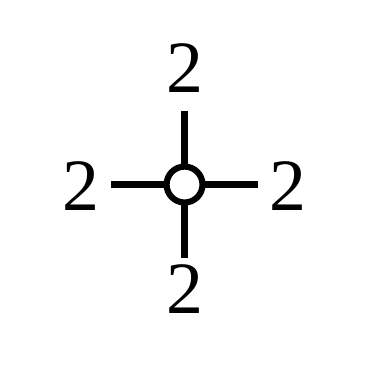}}=-h_i$, and $-\infty$ for all other tensor elements.
In cases where the local field vanishes, these dots reduce to the copy tensor in terms of the tropical algebra which demands that all the legs have the same indices. 
Contraction of the tensor network under the tropical algebra gives the ground state energy of the Ising spin glass. In the contraction, the $\oplus$ operator selects the optimal spin configuration, and the $\odot$ operator sums the energy contribution from subregions of the graph. The intermediate tensors record the minimal energy given the external tensor indices, so they correspond to max-marginals in the graphical model~\cite{mezard2009information}.

From a physics perspective, the tropical tensor network naturally arises from computing the zero-temperature limit of the partition function $Z =  \sum_{\{\sigma\} } e^{-\beta E} $. The ground state energy,  $E^{\ast} = -\lim_{\beta \rightarrow \infty}\frac{1}{\beta } \ln Z = -\lim_{\beta \rightarrow \infty} \frac{1}{\beta }\ln\sum_{\{\sigma\} } \prod_{i<j}e^{\beta J_{ij} \sigma_i \sigma_j}\prod_{i}e^{\beta h_{i} \sigma_i} $, involves ordinary sum and product operations for the Boltzmann weights. When taking the zero temperature limit, it is more convenient to deal with the exponents directly 
\begin{equation}
\lim_{\beta\rightarrow \infty}\frac{1}{\beta} \ln (e^{\beta x} + e^{\beta y}) = x \oplus y, \,\,\,\,\,\,\,\,\,\, \frac{1}{\beta}\ln ( e^{\beta x} \cdot e^{\beta y}) = x \odot y, \label{eq:lowTexpansion}
\end{equation}
which leads to the tropical algebra \Eq{eq:max-sum-alg}. The tropical representation also corresponds to the logarithmic number system~\cite{Kingsbury1971} which avoids the numerical issue in dealing with exponentially large numbers on computers with finite precision numerics~\cite{Zhu2019b}. 

Moreover, one can also employ the present approach to count the number of ground states at the same computational complexity of computing the ground state energy. To implement this, we further generalize the tensor element to be a tuple $(x, n)$ composed by a tropical number $x$ and an ordinary number $n$. The tropical number records the negative energy, while the ordinary number counts the number of minimal energy configurations. For tensor network contraction, we need the multiplication and addition of the tuple: $(x_1, n_1) \odot (x_2,n_2) = (x_1 + x_2, n_1\cdot n_2)$ and 
\begin{equation}
    (x_1, n_1)\oplus (x_2, n_2) = \begin{cases}
 (x_1\oplus x_2, \, n_1 + n_2 ) & \text{if $x_1 = x_2$} \\
 (x_1\oplus x_2,\, n_1 ) & \text{if $x_1>x_2$} \\
 (x_1\oplus x_2,\, n_2 )& \text{if $x_1 < x_2$}
 \end{cases}.
\end{equation}

Essentially, these two numbers in the tuple correspond to leading order and the $\mathcal{O}(1/\beta)$ contributions (energy and entropy) in the low-temperature expansion of the log-partition function. After contracting the tensor network, one reads out the ground state energy and degeneracy from the two elements of the tuple. In this way, one can count the number of optimal solutions exactly without explicitly enumerating the solutions~\cite{Zhang2009, Marinescu2019}.

\section{Contract Tropical Tensor Networks}

We have formulated the computation of the ground state energy and the ground state degeneracy of the Ising spin glass \Eq{eq:problem} as a contraction of the tropical tensor network. On a tree graph, contraction of the tropical tensor network is equivalent to the max-sum algorithm~\cite{koller2009probabilistic}, i.e. the maximum of a posterior version of the sum-product (belief propagation) algorithm on graphical models. 
On a general graph, when the junction tree algorithm~\cite{lauritzen1988local} applies it can be treated as a special case of the tropical tensor network contraction algorithm using a specific contraction order utilizing a tree decomposition of the graph.

The contraction of a general tensor network belongs to the class of \#P hard problems~\cite{Schuch2007}, so it is unlikely to find polynomial algorithms for exact contractions. 
Algorithmically, the computational complexity of tensor network contraction is exponential to the tree-width of the network~\cite{treewidth}. On a regular graph (e.g. 2D lattice), one can easily find a good contraction order that has an optimal computational complexity. 
However, on a general graph, a good contraction order is usually difficult to find, thus one usually relies on heuristic algorithms to identify a contraction order with low computational complexity. Ref.~\cite{treewidth} proposed to use tree decomposition of the line graph of the tensor network, found by a branch and bound algorithm. This has been widely adopted in subsequent works on classical simulation of quantum circuits with tensor networks~\cite{Boixo2017, Pednault, Fried2018, Dumitrescu2018a, Dudek2019, Villalonga, Schindler2020a, Schutski2020}. 
Recently, more advanced heuristic algorithms have been developed by combining graph partition algorithms and greedy algorithms~\cite{Gray2020, Huang2020b}.

In addition to a good contraction order, efficient linear algebra libraries are also important for the performance of the contractions. For ordinary contractions, the basic linear algebra subprograms (BLAS) library is a standard tool for performing efficient product and plus operations, and can fully release the computational power of specialized hardware such as GPUs and tensor processing units. 
For the tropical algebra, fortunately, basic operations can be inherited from standard linear algebra libraries as long as they are programmed in a generic manner to support $\oplus$ and $\odot$ operators.
When performing contractions on GPUs, another important factor is memory efficiency, that is, all operations should be performed in-place without allocating extra memory. This actually shares the same demand as the simulation of quantum circuits. To this end, one can actually contract tropical tensor networks by repurposing software that was originally developed for quantum circuit simulations. 

To sum up, the tropical tensor network formulation opens a way to leverage recent algorithmic and software advances in tensor network contraction for combinatorial optimization problems. Moreover, the tensor contraction formation fits nicely to the specialized hardware such as GPUs, where, as we reported below, one can actually employ low precision floating numbers (or even integer type for integral couplings) for better numerical performance and reduced memory usage. 



\section{Obtaining the Ground States with Automatic Differentiation } 
Given the way to compute the ground state energy of the spin glass, there are several ways to obtain the ground state configurations. The most straightforward way would be running the same energy minimization program repeatedly with perturbed fields. Since the ground state energy is a piecewise linear function of the fields, the numerical finite-difference of the energy with respect to fields suffices to determine the ground state configurations~\footnote{In cases of the degenerated ground state, the approach gives one out of many ground state configurations. The particular configuration is selected by the default implementation of the maximum function, which returns the first argument when the two arguments are equal. One could obtain other degenerate solutions by changing this default behavior.}. Alternatively, one can impose an arbitrary order of the spin variables and compute the conditional probability of a variable being in the ground state given the previous ones, 
then sample the ground state configurations according to the conditional probability~\cite{mezard2009information}. 
Both methods need to re-run the contraction algorithm $\mathcal{O}(N)$ times 
with the same memory cost as finding the ground state energy. 
One can nevertheless trade memory for computation time by caching intermediate contraction results and backtracking the computation for minimal energy configuration.

We employ the differentiable programming technique to differentiate through the tropical tensor network contraction~\cite{Liao2019a}. To this end, we program the whole tensor network contraction in a differentiable way and compute the gradient of the contraction outcome with respect to the tensor elements using automatic differentiation. We note that the general idea of differentiating through combinatorial optimization solver applies to cases beyond tropical tensor network contraction~\cite{blog}. It is well known that there is a time-space trade-off in different ways of performing the automatic differentiation to a computer program~\cite{Baydin2018}. The forward mode automatic differentiation (such as \texttt{ForwardDiff.jl}~\cite{Revels2016}) has the same time and memory cost as the finite difference approach. 
While in the other extreme limit, the reverse mode automatic differentiation (such as (\texttt{Nilang.jl}~\cite{Liu2020g}) displays the $\mathcal{O}(1)$ computation overhead compared to the forward tensor contraction, and $\mathcal{O}(N)$ memory overhead. The time versus memory trade-off can be further controlled flexibly by using the checkpointing technique~~\cite{Baydin2018}.

\section{Applications}
We first apply the tropical tensor network approach to the Ising spin glasses on $L\times L$ square lattices, with the tensor network shown in Fig.~\ref{fig:square}(a). 
Interestingly, the computation of tensor network contraction is similar to evolving a quantum state under the action of local quantum gates, with the crucial difference that we are now dealing with nonunitary gates with the tropical algebra. 

As shown in Fig.~\ref{fig:square}(b), the tensor network is cast into the expectation of a tropical circuit on the state vector of $2^L$ dimension. We denote $\raisebox{-1ex}{\includegraphics[scale=0.25, trim={0cm 0.8cm 0.8cm 0cm}, clip]{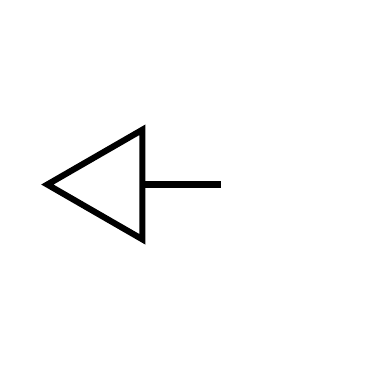}} = \left(\begin{matrix}
        0 \\
        0
    \end{matrix}\right)$ so that the initial and final states are both product state 
    $\left(\begin{matrix}
        0 \\
        0
    \end{matrix}\right)^{\otimes L}$. The square symbols represent tropical gates, in which $
    \raisebox{-2.4ex}{\includegraphics[scale=0.25, trim={0cm 0.0cm 0cm 0cm}, clip]{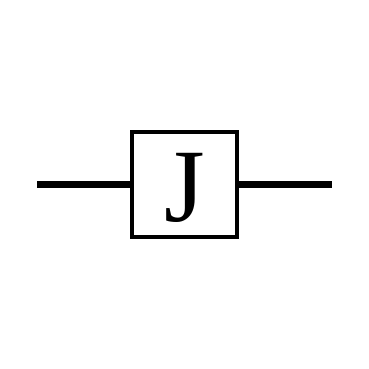}} = \left(\begin{array}{cc} J_{ij} & -J_{ij} \\  -J_{ij} & J_{ij} \end{array}\right) $ and $
    \raisebox{-2.4ex}{\includegraphics[scale=0.25, trim={0cm 0.0cm 0cm 0cm}, clip]{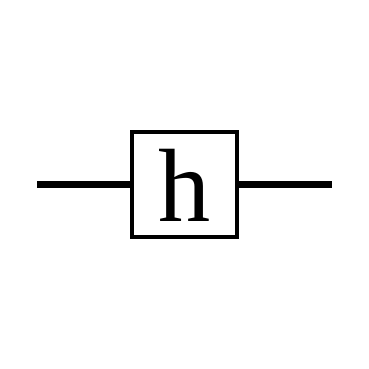}} = \left(\begin{array}{cc} h_{i} & -\infty \\  -\infty & -h_i \end{array}\right)  $ are single-site gates. The symbol $ 
        \raisebox{-3.8ex}{\includegraphics[scale=0.4, trim={0cm 0.0cm 0cm 0cm}, clip]{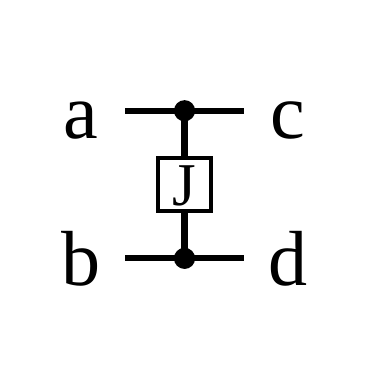}}$ denotes two site gates acting on neighboring sites. 
        In fact, it is a diagonal tropical matrix $\mathrm{diag}(J_{ij}, -J_{ij}, -J_{ij}, J_{ij})_{ab,cd}$, with the off-diagonal elements set to $ -\infty$. The order of operation of these diagonal gates to the state vector can be arbitrary.  
\begin{figure}[t]
\centering
    \includegraphics[width=\columnwidth]{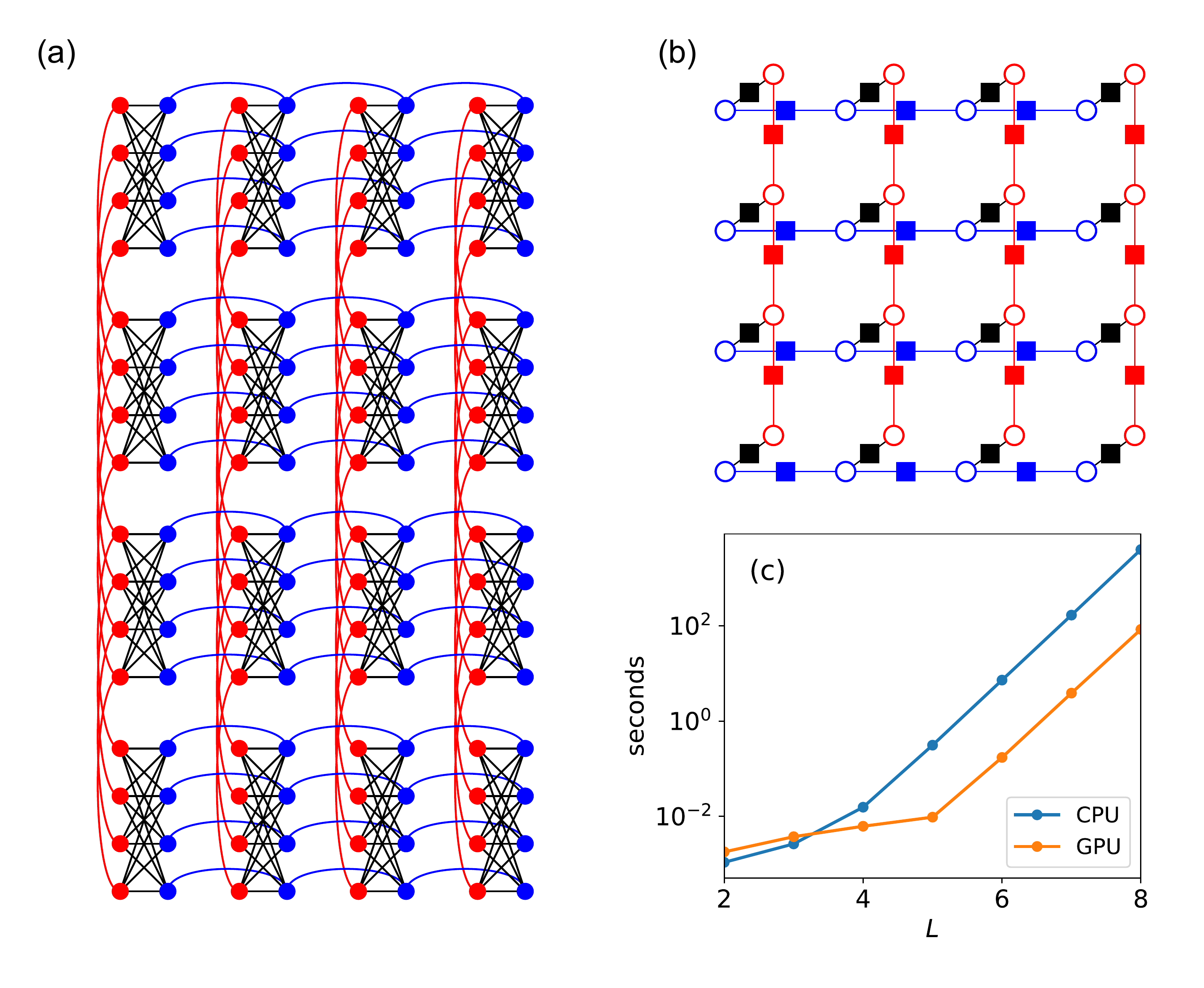}
    \caption{(a) A chimera lattice with $4\times 4$ unit cells. Dots represent Ising spins and lines indicate couplings. (b) Tensor network representation, where each node has a degree of freedom of 4 spins. (c) Wall clock time for computing the ground state energy of Ising spin glass on the chimera graph with the $L\times L$  unit cell ($8L^2$ spins).
}  
 \label{fig:chimera}
\end{figure}
Exploiting this intimate connection, 
we employ the quantum programming software \texttt{Yao.jl}~\cite{Luo2020} to contract these tropical tensor networks~\footnote{In general, it is always possible to map the tensor network contraction to a quantum circuit simulation by possibly introducing extra ancilla qubits.}.
It enables us to obtain the ground state energy of $1024$ spins with external fields in about 590 seconds on a single Nvidia V100 GPU, with single-precision floating numbers \texttt{Float32} for the tensor elements.

Next, we consider spin glass instances with $\pm J$ coupling and no external field on the chimera graph of the actual D-Wave device~\cite{Johnson2011} shown in Fig.~\ref{fig:chimera}(a). The chimera graph consists of unit cells arranged in a square grid of the size of $L\times L$. Each unit cell contains 8 spins forming a complete bipartite graph. Each group of four spins within the unit cell connects horizontally or vertically to the spins in the neighboring unit cells. 
\begin{figure}[t]
\centering
    \includegraphics[width=\columnwidth]{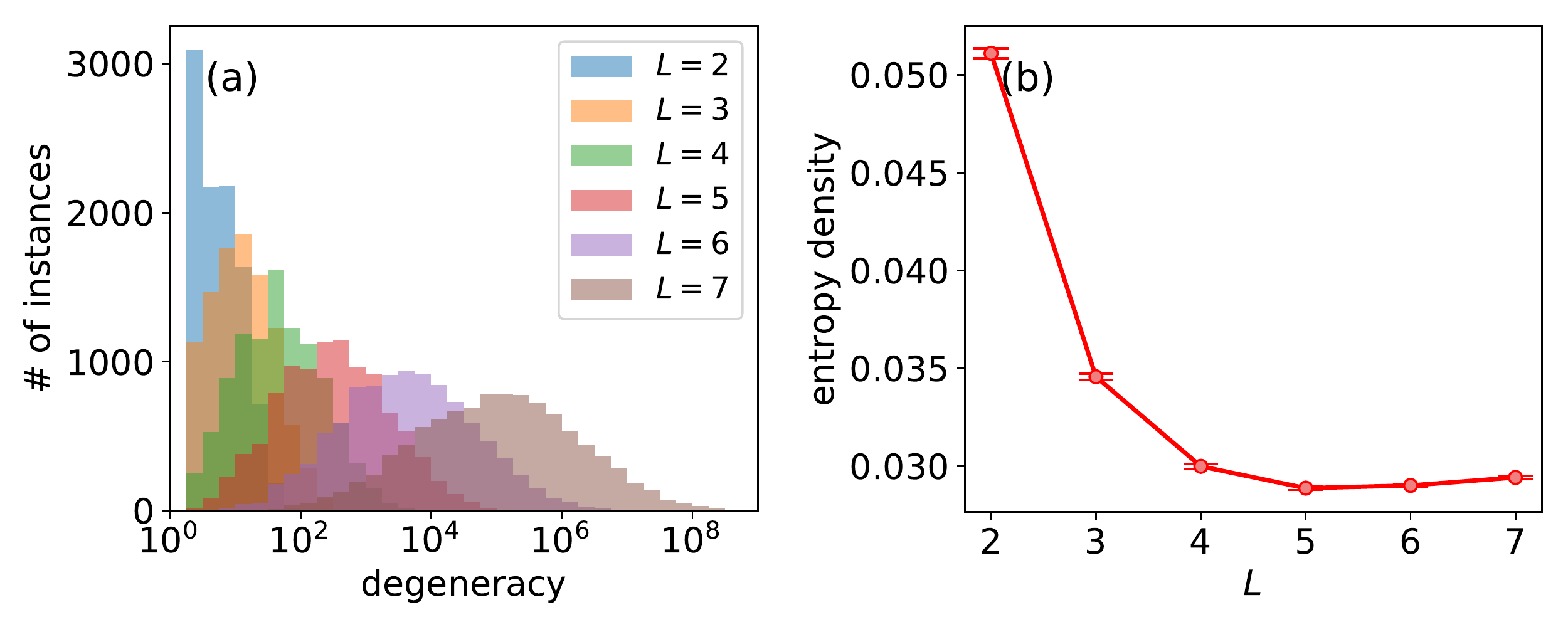}
\caption{(a) Histogram of the ground state degeneracy of $\pm J$ spin glasses on the chimera graph with $L\times L$ unit cells ($8L^2$ Ising spins). For each system size, we solve $10000$ random instances.  (b) The residual entropy density versus system size. 
} \label{fig:degeneracy}
\end{figure}
We transform the chimera graph into a tensor network shown in Fig.~\ref{fig:chimera}(b) by exploiting its specific structure~\cite{Selby2014}. The red and blue circles are tropical copy tensors that represent a group of four Ising spins within each unit cell. The black tensor describes the intra-unit-cell couplings. While the red and blue squares denote the intercell interaction in the vertical and horizontal direction respectively. These tensors are all $16 \times 16$ tropical matrices that contain the couplings between the original Ising spins. Such a tensor network formulation makes better use of the bipartite structure of the chimera graph than simply grouping the $8$ spins within the unit cell together~\cite{Rams2018a}. After turning these tensors into local tropical gates, contraction of the tensor network can be carried out as evolution of a state with dimension $16^{L}$. As shown in Fig.~\ref{fig:chimera}(c) one can obtain the ground state energy of $8L^2 = 512$ Ising spins in $84$ seconds on the Nvidia V100 GPU. This is much faster than brute force enumeration using GPUs~\cite{bruteforce}. It is also slightly faster than the belief propagation exact solver running on 16 CPU cores used in Ref.~\cite{Boixo2014}. We use \texttt{Int16} data type for computational and memory efficiency, which is sufficient for such calculation since the energy has bounded integral values. 

Figure~\ref{fig:degeneracy}(a) shows the histogram of the ground state degeneracy of the chimera spin glasses. One observes that the distributions are unimodal and broaden as the system size enlarges. Figure~\ref{fig:degeneracy}(b) shows the residual entropy density $s = \mathbb{E}[\ln g ]/ (8L^2)$  where $g$ is the degeneracy and the expectation is over the $10000$ random instances. The value of the residual entropy approaches $s=0.03$ for increasingly larger system sizes. As a comparison, this value of the entropy density is smaller than the one of the $\pm J$ square lattice Ising spin glass $s\approx0.07$~\cite{Vannimenus1977,Morgenstern1980,Cheung1983,Wang1988,PhysRevLett.69.2292,PhysRevE.48.R3221}, indicating a smaller number of degenerated ground state on the chimera graph compared to the  $\pm J$ square lattice spin glasses, possibly due to the larger connectivity in the Chimera graph which induces more constraints to each spin in the ground-state and suppresses the degeneracy.

For problems on more general graphs, our method benefits from the contraction order developed in the quantum computation community~\cite{boixo2017simulation,boixo2018characterizing,kourtis2019fast,Gray2020,Huang2020b}. 
As an example, with the present approach one can compute optimal solutions and count the number of solutions for spin glasses and combinatorial optimization problems on random graphs with hundreds of nodes, and check numerically the replica symmetry mean-field solutions~\cite{Mezard2001,Mezard2003}. Details can be found at the Appendix.

\section{Discussions}

An immediate implication of our method is that quantum circuit simulators can be repurposed to solve combinatorial optimization problems. This connection adds a profitable motivation for crafting efficient and generic quantum circuit simulators besides validating quantum devices. 

We notice that the state-of-the-art method branch-and-cut approaches are able to achieve better performance for spin glasses on 2D lattices. For example, Ref.~\cite{de1995exact} reached $100\times 100$ lattices for a spin glass with Gaussian couplings, and $50\times 50$ lattices with $\pm$J couplings~\cite{de1996exact}. However, the branch and bound method is less efficient in computing degeneracies. For example, the branch-and-bound results for entropy were reported with for $8\times 8$ lattices~\cite{percus2006computational}, while, our method works out the ground-state entropy of $\pm J$ spin glass on $32\times 32$ lattices.
Moreover, the linear programming bounding method is sensitive to coupling types and connectivity of the model. On $2$D lattices, the branch-and-cut method is quite efficient when equipped with the \textit{circle inequality}~\cite{de1995exact} technique, especially with Gaussian couplings. But it turns out to be less efficient when the topology is a $3$D lattice, where only results with $ 4 \times 4\times 4=64$ spins are reported in the literatures~\cite{percus2006computational}. In contrast, on $3D$ lattices, our method works to $6\times 6\times 6=216$ spins.
More seriously, if the model changes from an Ising spin glass to a Potts glass, not only the cutting plane method but also the linear programming bounding method breaks down. As a relief, one has to develop a more sophisticated Semi-Definite Programming (SDP) method for providing energy lower bounds~\cite{ghaddar2011branch,anjos2013solving}. Reference~\cite{ghaddar2011branch} computed the ground-state energy of a $\pm J$ $3$-state Potts glass model on a $9\times 9$ lattice using $10$ hours. As a comparison, our method is able to compute both ground-state energy and entropy on $18\times 18$ lattices in $10$ minutes, thus is significantly superior to SDP based branch-and-cut methods for Potts models (see \App{app:potts}). Moreover, one could also apply specific bounds on the ground-state energy to enforce sparsity of the tropical tensors, this would combine the tropical tensor network framework with the branch and bound methods. 

Moving forward, approximated contraction schemes for the tropical tensor networks may provide practical algorithms for the optimization and counting of large instances. A \textsc{Julia} implementation of the tropical tensor network used in this paper is available at Ref.~\cite{github}. Thanks to generic programming, a minimalist working example contains only $\sim60$ lines of code. 

\begin{acknowledgments}
    We thank Hai-Jun Liao, Zhi-Yuan Xie, and the BFS Tensor community for inspiring discussions, and Yingbo Ma for discussions on the Tropical BLAS library~\cite{ma}.
    P.Z. is supported by projects QYZDB-SSW-SYS032 of CAS, and Project 12047503 and 11975294 of NSFC.  L.W. is supported by the National Natural Science Foundation of China under Grant No.~11774398, and the Ministry of Science and Technology of China under Grant No.~2016YFA0300603 and No.~2016YFA0302400.
\end{acknowledgments}

\bibliography{refs.bib}

\appendix

\section{Mapping a tensor network to a quantum circuit}\label{app:mapping}

We first introduce notations that used in representing a tropical circuit.

\begin{enumerate}
    \item Starting/termination symbol

\begin{align}
    \raisebox{-1ex}{\includegraphics[scale=0.25, trim={0cm 0.8cm 0.8cm 0cm}, clip]{sym_tri.pdf}} = ~\left(\begin{matrix}
        0 \\
        0
    \end{matrix}\right)
\end{align}

    \item Horizontal coupling gate

\begin{align}
    \raisebox{-2.4ex}{\includegraphics[scale=0.25, trim={0cm 0.0cm 0cm 0cm}, clip]{sym_horizontal_J.pdf}} = ~\left(\begin{array}{cc} J_{ij} & -J_{ij} \\  -J_{ij} & J_{ij} \end{array}\right)
\end{align}

    \item Magnetic field gate
\begin{align}
    \raisebox{-2.4ex}{\includegraphics[scale=0.25, trim={0cm 0.0cm 0cm 0cm}, clip]{sym_horizontal_h.pdf}} = ~\left(\begin{array}{cc} h_{i} & -\infty \\  -\infty & -h_i \end{array}\right)
\end{align}

    \item Vertical coupling gate

\begin{align}
        \raisebox{-3.0ex}{\includegraphics[scale=0.3, trim={0cm 0.0cm 0cm 0cm}, clip]{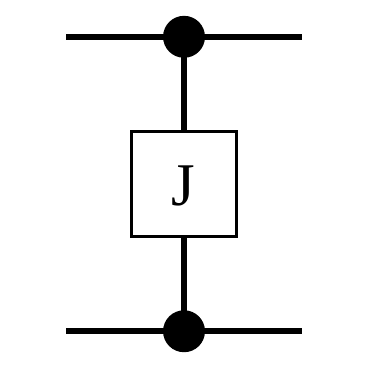}} = ~\left(\begin{array}{cccc}
            J_{ij} & -\infty & -\infty & -\infty\\
            -\infty & -J_{ij} & -\infty & -\infty\\
            -\infty & -\infty & -J_{ij} & -\infty\\
            -\infty & -\infty & -\infty & J_{ij}
        \end{array}\right)
\end{align}

    \item Copy gate
\begin{align}
    \raisebox{-2.5ex}{\includegraphics[scale=0.25, trim={0cm 0.0cm 0cm 0cm}, clip]{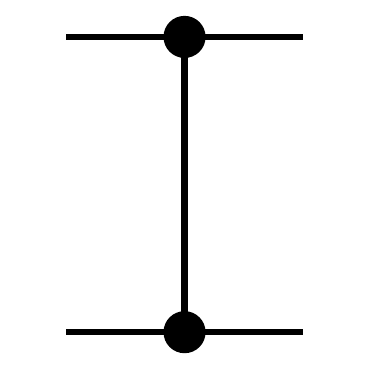}} ~=~ \left(\begin{matrix}
        0 & -\infty & -\infty & -\infty\\
        -\infty & -\infty & -\infty & -\infty\\
        -\infty & -\infty & -\infty & -\infty\\
        -\infty & -\infty & -\infty & 0
    \end{matrix}\right).
    \label{eq:symbols-copy}
\end{align}


    \item Cut gate
\begin{align}
    \raisebox{-2.5ex}{\includegraphics[scale=0.25, trim={0cm 0.0cm 0cm 0cm}, clip]{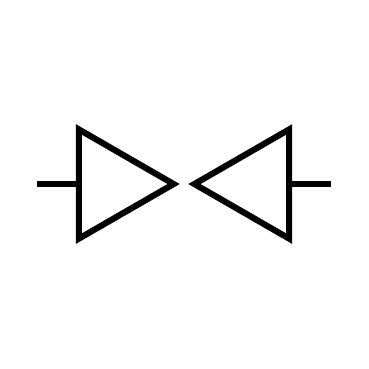}} ~=~ \left(\begin{matrix}
        0 & 0\\
        0 & 0
    \end{matrix}\right).
    \label{eq:symbols-cut}
\end{align}

\end{enumerate}

\begin{figure}[t]
\centering
    \includegraphics[width=0.9\columnwidth, trim={0 3cm 0 0cm}, clip]{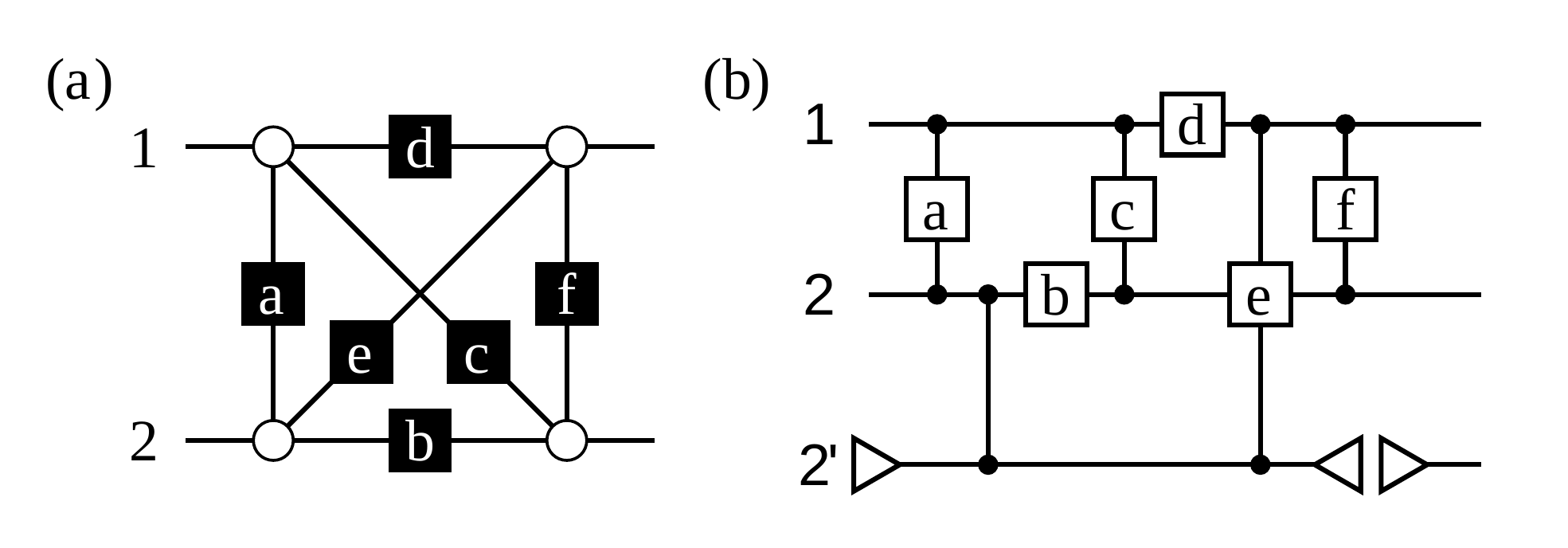}
    \caption{Mapping the tensor network in (a) to a tropical quantum circuit in (b).} \label{fig:mapping}
\end{figure}

The copy gate and cut gate are useful in mapping a general tropical tensor network to the circuit model. As an example, in \Fig{fig:mapping} (a), in order to arrange gates in specific time order, we introduce an extra ancilla qubit \texttt{2\textquotesingle} as shown in (b).
One can use the copy gate to store the information in qubit \texttt{2} into the ancilla qubits \texttt{2\textquotesingle}.
At the end of an operation, we use the cut gate to restore the state of the ancilla qubit.

\section{Reversible programming approach to compute gradients}\label{app:nilang}

It is a challenge to differentiate a generic quantum simulator with tropical numbers inside.
We need to derive the backward rules for tropical quantum circuits simulation.
Unlike a traditional quantum simulation program, one can not trace back the intermediate states by applying the adjoint of gates to save memory~\cite{Luo2020}.

\begin{figure}
\centering
    \includegraphics[width=0.96\columnwidth,trim={0 4cm 0 0},clip]{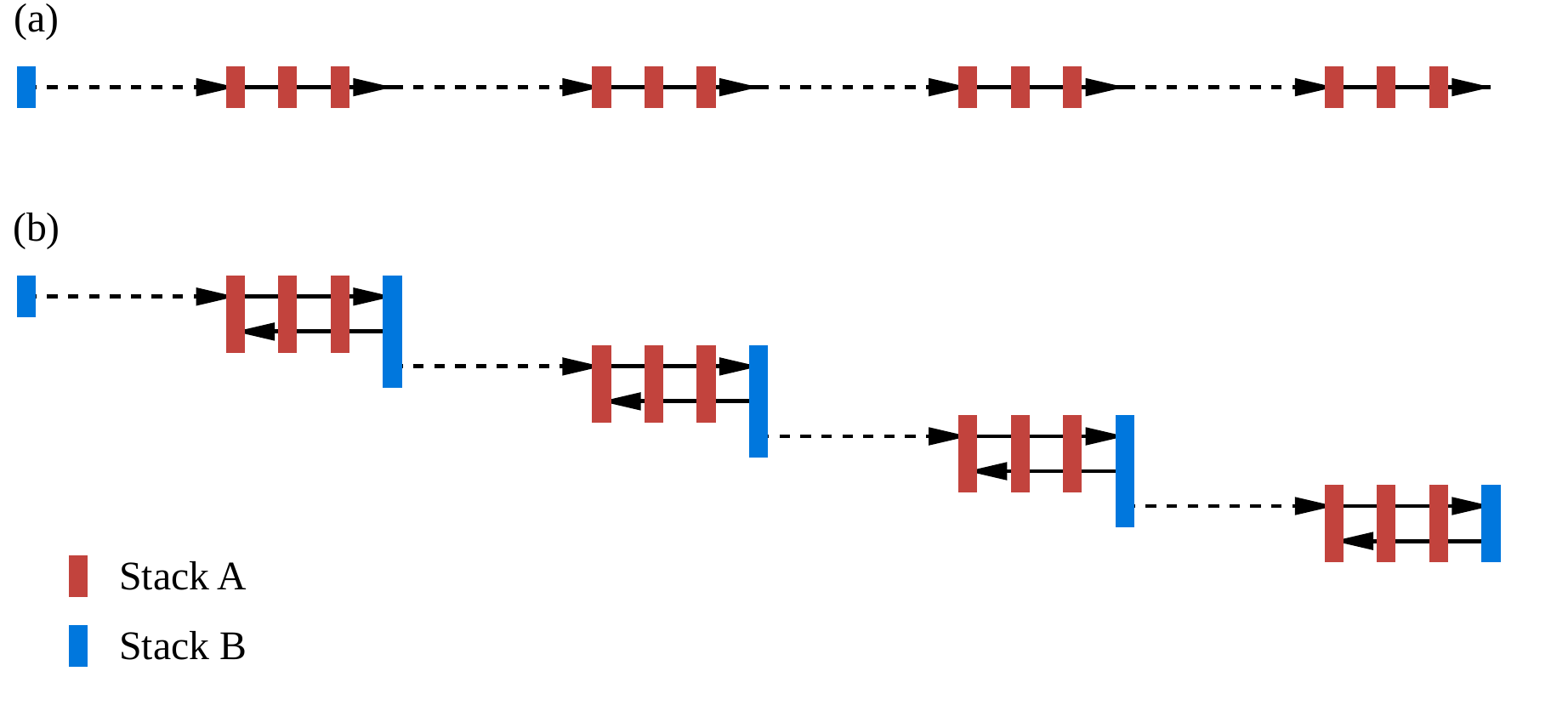}
    \caption{The compute-copy-uncompute paradigm in reversible programming. Rectangles represent memory allocation. Dashed lines are reversible operations (e.g. the vertical coupling gates, magnetic field gate, and copy gate), while the solid lines represent operations that require caching intermediate states to keep it reversible. (a) the naive algorithm that caches states every step, (b) the algorithm that uncomputes stack \texttt{A} after sweeping each column.} \label{fig:uncomputing}
\end{figure}

Instead of deriving the backward rule manually, we differentiate the source codes by writing it in a reversible programming manner~\cite{Liu2020g}.
Due to the overhead of reversible programming, the memory usage of our reversible implementation is $2L$ times the original program, while the computational time is also several times slower. This overhead is acceptable in differential programming since it is comparable to the theoretical optimal of the checkpointing scheme in traditional machine learning.
In \Fig{fig:uncomputing}, we illustrate the compute-copy-uncompute scheme in reversible programming.
Figure~\ref{fig:uncomputing}(a) is the naive approach that caches all intermediate states in a global stack with a negligible computational time overhead.
It uses approximately $L^2$ times more memory than the original program.
Since the spin-glass solver is memory critical, a better approach is to uncompute some of the intermediate results as shown in \Fig{fig:uncomputing}(b).
In \Fig{fig:uncomputing}(b), we use two stacks. \texttt{A} stack is a dynamic one that uncomputed in each sweep of a column. \texttt{B} stack is a global one, that only uncomputed when running the program backward. Both \texttt{A} and \texttt{B} are $L$ times the size of a state vector, hence the memory overhead is $2L$ and the computational time overhead is $\sim 2$.

\begin{figure}
\centering
    \includegraphics[width=0.96\columnwidth,trim={0cm 1cm 10cm 0cm},clip]{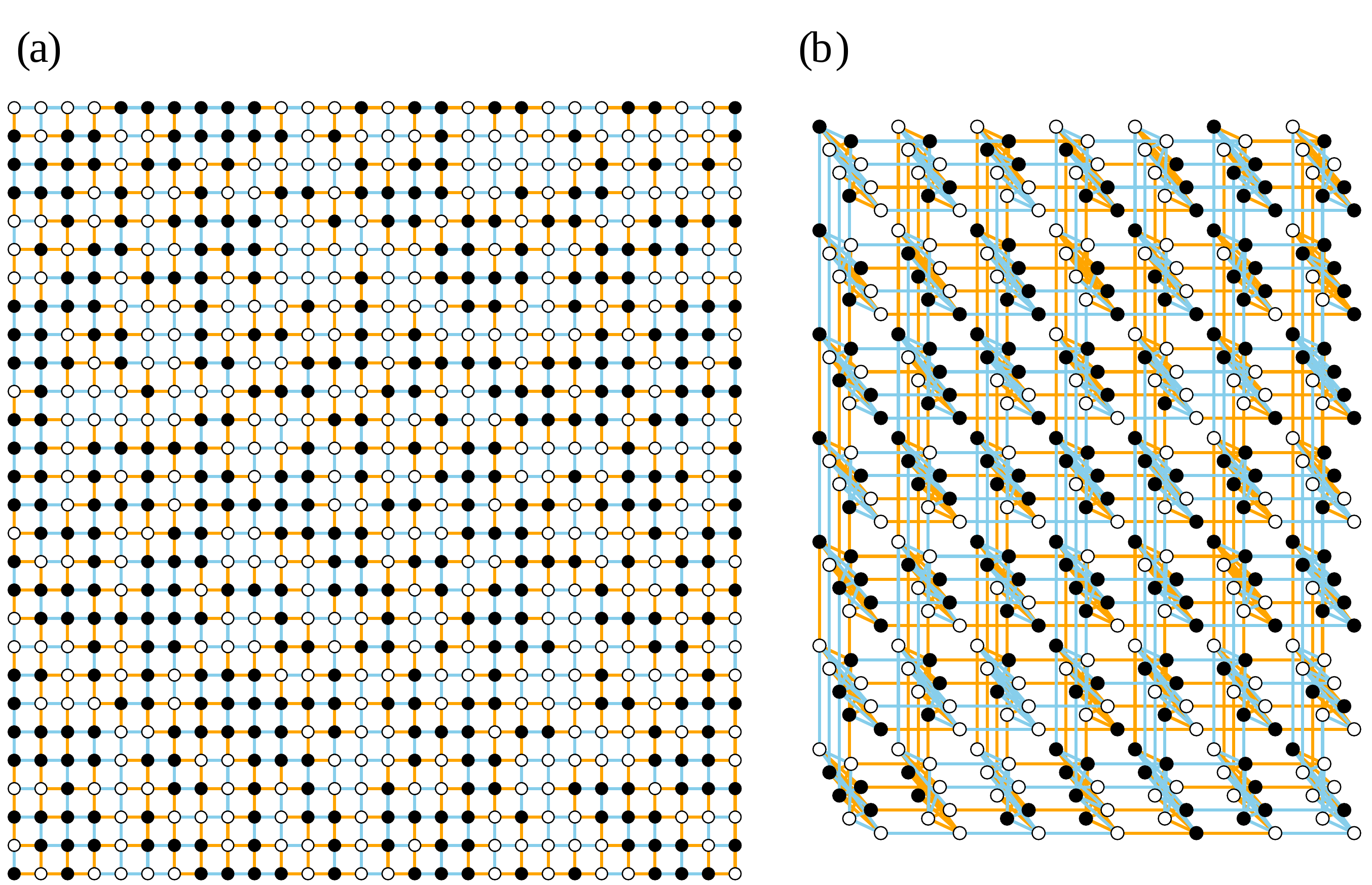}
    \caption{(a) $28 \times 28$ square lattice Ising spin glass with an optimal configuration.
    (b) $7 \times 7$ Chimera lattice Ising spin glass with an optimal configuration. } \label{fig:spinglassconfig}
\end{figure}

\begin{figure}[t]
\centering
\includegraphics[width=\columnwidth]{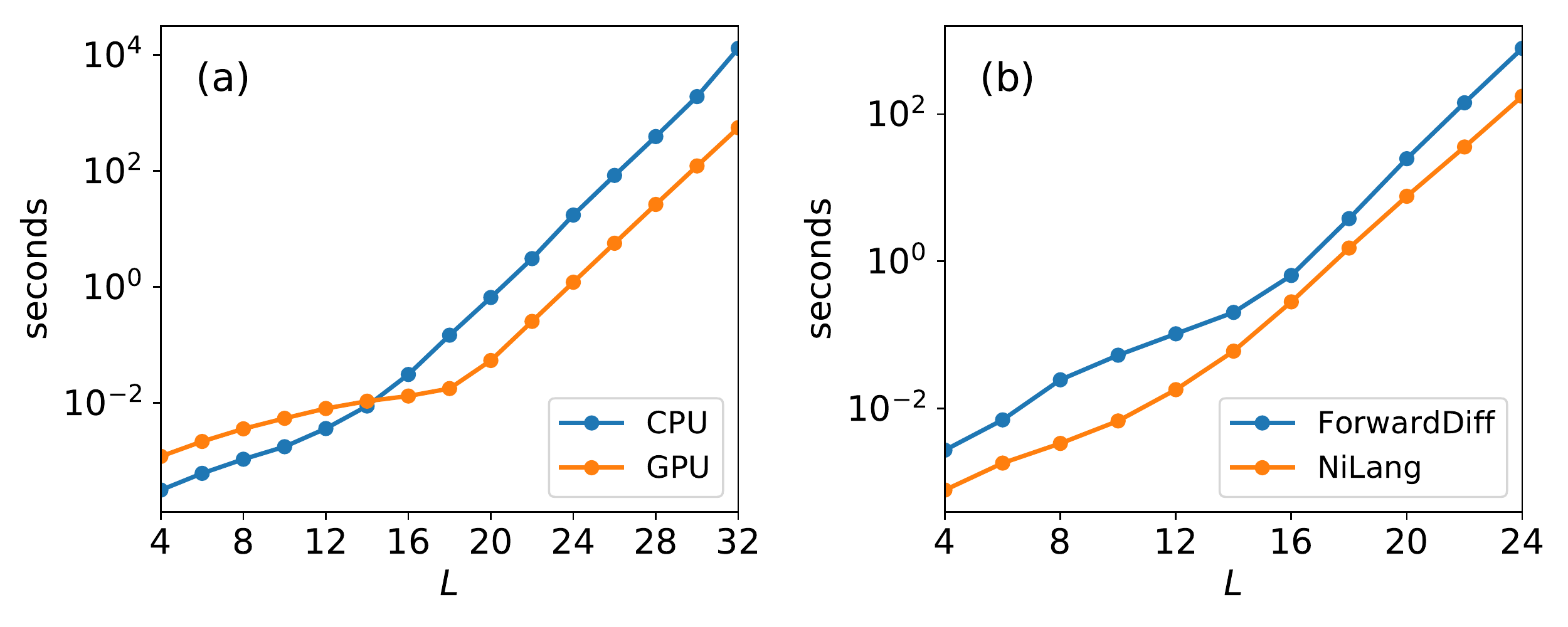}
\caption{Wall clock time for computing the ground state energy of the (a) Ising spin glass on an open square lattice with $L^2$ spins. (b) Wall clock time for computing the ground state configurations using forward (\texttt{ForwardDiff.jl}~\cite{Revels2016}) on GPU and reverse mode (\texttt{Nilang.jl}~\cite{Liu2020g}) automatic differentiation on CPU respectively.
\label{fig:performance}} 
\end{figure}

Figure~\ref{fig:performance}(a) shows the wallclock time for computing the ground state energy of Ising spin glass on the square lattice with Gaussian random couplings and fields. One can obtain the ground state energy of $1024$ spins with external fields in about 590 seconds on a single Nvidia V100 GPU, with single-precision floating numbers \texttt{Float32} for the tensor elements.
We further compared the performance of finding out the ground state configuration using the forward mode (\texttt{ForwardDiff.jl}~\cite{Revels2016}) and reverse mode (\texttt{Nilang.jl}~\cite{Liu2020g}) automatic differentiation respectively. 
The reverse mode automatic differentiation is more efficient in this application than the forward model AD which has computational complexity proportional to the number of parameters $L^2$. However, the reverse mode AD requires caching intermediate states for back-propagation, which causes memory overheads.
\texttt{NiLang.jl} provides machine instruction level automatic differentiation. One does not need to derive the backward rules manually, instead, he can just rewrite the original program in reversible programming~\cite{Perumalla2013} style and the automatic differentiation just works. Reversible programming also provides a flexible tradeoff between space and time, so that we can differentiate a spin-glass solver up to $L=28$ with an $\mathcal{O}(L)$ space overhead (see \App{app:nilang}). In \Fig{fig:performance} (b), we show the timings up to $L=24$.

We can see from the figure that, although with non-negligible overhead, the reversible programming approach is still much more efficient than the forward mode AD since the computational overhead of forward-mode AD is proportional to the number of parameters $L^2$. In the benchmark shown in the figure, even single thread reversible programming AD is faster than forward-mode AD on GPU by a factor of $\sim 6$.
In \Fig{fig:spinglassconfig}, we show the optimal configuration of Ising spin-glass models on a $28 \times 28$ square lattice and a $7 \times 7$ chimera lattice.

%

\section{Ground-state energy and entropy for Potts spin glasses on square lattice}\label{app:potts}
\begin{figure}[h]
\centering
\includegraphics[width=0.6\columnwidth]{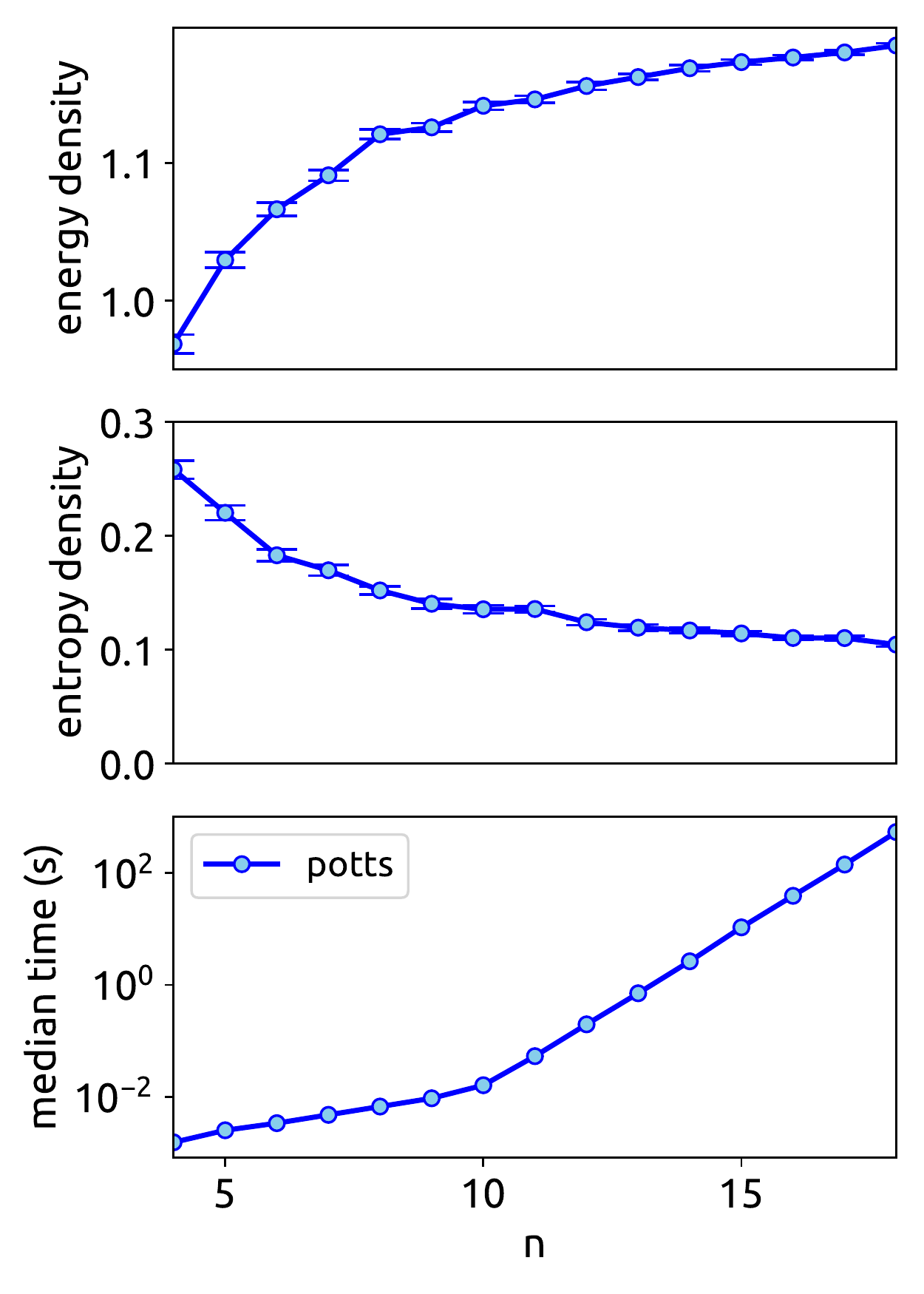}
    \caption{Ground-state energy, entropy,,,,,,,,, and computational time of $q=3$ state Potts spin glass model(with Hamiltonian defined in~\cite{ghaddar2011branch}) on square lattices.
        Each data point is averaged over $100$ random instances computed on a single GPU.
        As a comparison, the existing branch-and-cut method with the semi-definition programming energy lower bounds method~\cite{ghaddar2011branch} on the same model works up to $9\times 9$ lattices (using $10$ hours).
        \label{fig:potts}.}
\end{figure}

We notice that if the model changes from Ising spin glass to Potts glass, the branch-and-cut methods are not efficient. Indeed, not only the cutting plane method, but also the linear programming bounding method breaks down. As a relief, one has to develop more sophisticated Semi-Definite Programming (SDP) method for providing energy lower bounds~\cite{ghaddar2011branch}~\cite{anjos2013solving}. In the literatures, even with SDP bounding, one can only deal with $\pm J$ $3$-state Potts glass model on $9\times 9$ lattice, taking $10$ hours (see Tab.5.5 of ~\cite{ghaddar2011branch}). In contrast, our method is able to compute both ground-state energy and entropy on $18\times 18$ lattices in several minutes, thus is significantly superior to SDP based branch-and-cut methods for Potts models. The computational time is shown in Fig.~\ref{fig:potts}, where the Hamiltonian is defined as~\cite{ghaddar2011branch}
$$H = \sum\limits_{\langle i,j\rangle}J\left(\begin{matrix}1 & -1/2 & -1/2 \\ -1/2 & 1 & -1/2 \\ -1/2 & -1/2 & 1\end{matrix}\right)_{s_i,s_j},$$ where $s_i,s_j \in \{1,2,3\}$. 

\section{Counting number of optimal solutions in spin glasses and max 2-SAT problem on random graphs}\label{app:2sat}
\begin{figure}[h]
\centering
\includegraphics[width=0.6\columnwidth]{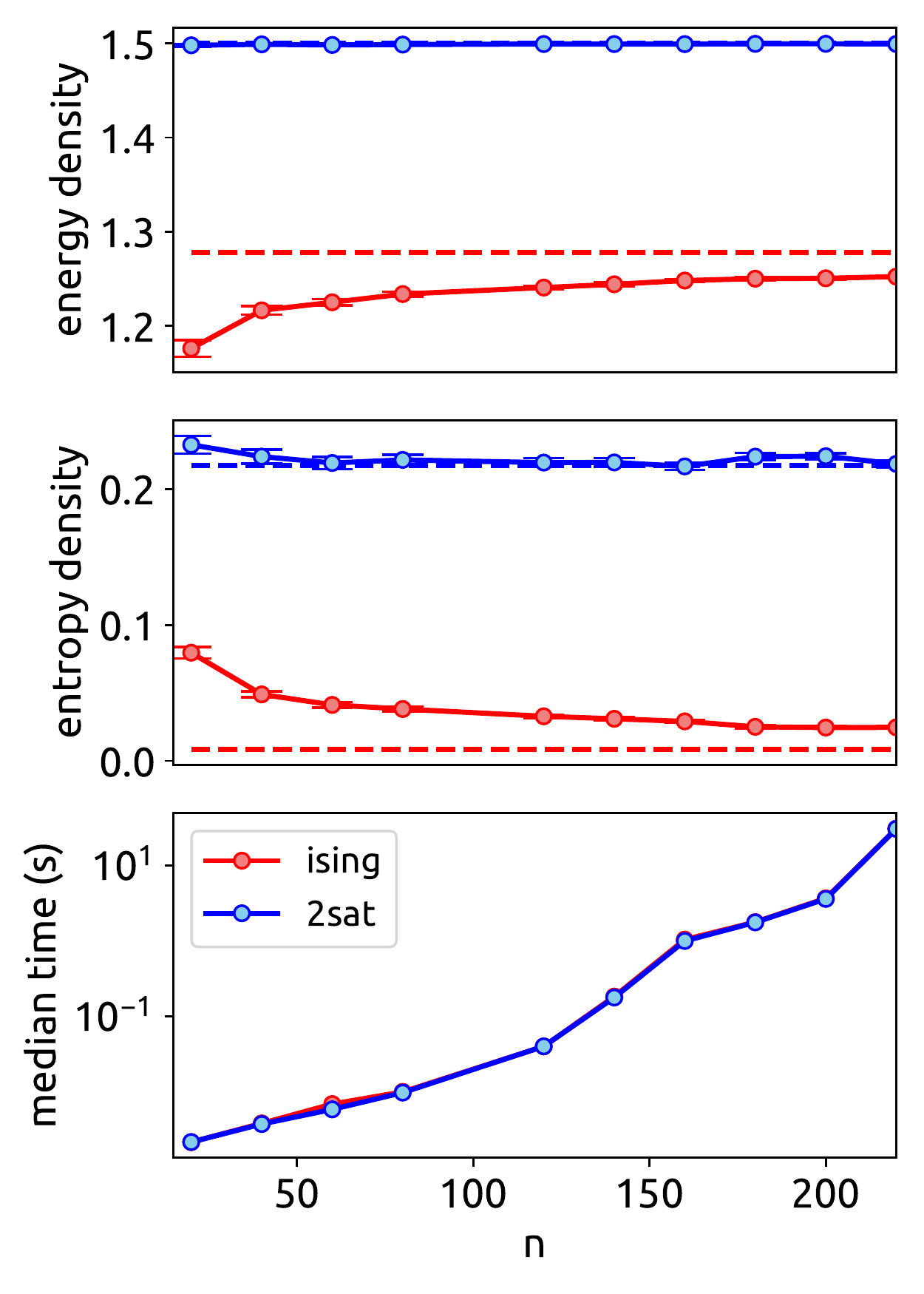}
    \caption{Ground-state energy and entropy of $\pm J$ spin glasses and MAX $2$-SAT problem on regular random graphs with degree $3$. Each data point is averaged over $100$ random instances, and were computed on a single GPU. The dashed lines are replica symmetric mean-field solutions using the cavity method~\cite{Mezard2001,Mezard2003}.\label{fig:rer}}
\end{figure}

Our method benefits from the fast-developing field of contraction order~\cite{boixo2017simulation,boixo2018characterizing,Kourtis2019,Gray2020,Huang2020b} approaches developed in the quantum computation community. This extends the ability of our approach from computing spin glasses on lattices to arbitrary graphs. We take the spin glasses and counting of combinatorial optimization problems on random graphs as an example.
The state-of-the-art method~\cite{Kourtis2019} for counting number of solutions for the Constraint Satisfaction Problems (CSP) are based on standard tensor network methods with 
enhanced contraction order. However it only works when the optimal solution is known, that is, all constraints can be satisfied and does not
work when the ground-state energy is unknown. Our method works not only for the CSP, but also for counting of optimization problems whose optimal solution needs to be determined first before counting the number of them. 
As an example to demonstrate the superiority of our method, we take the $\pm$ J spin glasses and \# Max-2-SAT problem on $3$ regular random graphs. These two problems are two distinct problems, whose counting problems all belong to the \#-P problem and no efficient exact algorithm exist. The results are plotted in Fig.~\ref{fig:rer}. From the figure, we can see that 
\begin{itemize}
\item the computational time for solving either $\pm$ J Ising or Max-2SAT are exactly the same, because our approach is general to treat all optimization and counting problems defined on the same graph, with exactly the same computational complexity.
\item Our method significantly better performance than the previous fast counting methods~\cite{Kourtis2019}. For example, on $3$ regular random graphs, the method of ~\cite{Kourtis2019} needs $100$ seconds, while our method takes only less than $10$ seconds, despite the fact that the problem we solved are much harder than that of ~\cite{Kourtis2019}.
\item Our exact results on both ground-state energy and entropy of the 2SAT problem coincide very well with the replica symmetry solution computed using the cavity method in~\cite{Mezard2001,Mezard2003}. However, the results of $\pm J$ Ising spin glass deviate significantly from the replica symmetry mean-field solution. This is actually not strange because the replica symmetry of $\pm J$ spin glass on $3$ regular random graphs is broken at low temperature, thus system is in the full replica symmetry breaking phase. Moreover this may also induce a large finite-size effect.
\end{itemize}

\end{document}